\documentclass[journal]{IEEEtran}

\usepackage{cite}
\usepackage{amsmath}
\usepackage{amssymb}
\usepackage{amsfonts}
\usepackage{graphicx}
\usepackage{textcomp}
\usepackage{xcolor}
\usepackage{array}
\usepackage{subfigure}
\usepackage[font=normalsize]{caption}
\usepackage{stfloats}
\usepackage{url}
\usepackage{verbatim}
\usepackage[export]{adjustbox}
\usepackage{ntheorem}
\usepackage{longtable}
\usepackage{booktabs}
\usepackage{makecell}
\usepackage{mwe}
\usepackage{graphbox}
\usepackage{enumitem}
\usepackage{ragged2e}
\usepackage{multicol} 
\usepackage[colorlinks, linkcolor=black, anchorcolor=black, citecolor=blue]{hyperref}
\usepackage{comment}
\usepackage{color}
\usepackage{bbding} 
\usepackage{diagbox} 
\usepackage{float}
\usepackage{multirow}
\theorembodyfont{\upshape}

\makeatletter
\renewtheoremstyle{plain}
{\item{\theorem@headerfont ##1\ ##2\theorem@separator}~}
{\item{\theorem@headerfont ##1\ ##2\ (##3)\theorem@separator}~}
\makeatother
\def\BibTeX{{\rm B\kern-.05em{\sc i\kern-.025em b}\kern-.08em
		T\kern-.1667em\lower.7ex\hbox{E}\kern-.125emX}}
\usepackage{balance}
\graphicspath{{QUDM/EPS/}}
\usepackage{lipsum}


\usepackage{setspace}
\graphicspath{{}}
\setenumerate[1]{itemsep=0pt,partopsep=0pt,parsep=\parskip,topsep=0pt}
\setitemize[1]{itemsep=0pt,partopsep=0pt,parsep=\parskip,topsep=0pt}
\setdescription{itemsep=0pt,partopsep=0pt,parsep=\parskip,topsep=0pt}

\usepackage{algorithm} 
\usepackage{algorithmicx}  
\usepackage{algpseudocode}  
\begin{document}
\title{QTP-Net: A Quantum Text Pre-training Network for Natural Language Processing}	
\author{Ren-Xin Zhao
		
\IEEEcompsocitemizethanks{
\IEEEcompsocthanksitem{
}
\IEEEcompsocthanksitem{

Ren-Xin Zhao is with the School of Computer Science and Engineering, Central South University, China, Changsha, 410083 and visiting the School of Computer Science and Statistics, Trinity College Dublin, Ireland, Dublin, D02 PN40 (Email: renxin\_zhao@alu.hdu.edu.cn). 
}
}
}

\maketitle
	
\begin{abstract}
Natural Language Processing (NLP) faces challenges in the ability to quickly model polysemous words. The Grover's Algorithm (GA) is expected to solve this problem but lacks adaptability. To address the above dilemma, a Quantum Text Pre-training Network (QTP-Net) is proposed to improve the performance of NLP tasks. First, a Quantum Enhanced Pre-training Feature Embedding (QEPFE) is developed to encode multiple meanings of words into quantum superposition states and exploit adaptive GA to fast capture rich text features. Subsequently, the QEPFE is combined with the Enhanced Representation through kNowledge IntEgration (ERNIE), a pre-trained language model proposed by Baidu, to construct QTP-Net, which is evaluated on Sentiment Classification (SC) and Word Sense Disambiguation (WSD) tasks. Experiments show that in SC, the QTP-Net improves the average accuracy by 0.024 and the F1 score by 0.029 on six benchmark datasets, comprehensively outperforming both classical and quantum-inspired models. In WSD, it reaches 0.784 average F1 score, which is 0.016 higher than the sub-optimal GlossBERT, and significantly leads on SE2, SE13, and SE15. QTP-Net provides a new solution for implicit semantic modeling in NLP and lays the foundation for future research on quantum-enhanced models.
\end{abstract}
	
\begin{IEEEkeywords}
natural language processing, adaptive Grover's algorithm, quantum computing, sentiment classification, word sense disambiguation
\end{IEEEkeywords}

\section{Introduction}\label{sec1}
NLP utilizes deep learning techniques to analyze and understand language, covering numerous branches such as SC for analyzing textual sentiment tendencies and WSD for determining object semantics based on context. For the SC, For SC, the HDL-Fuzzy-RMDL model achieves 92.3\% accuracy on the SemEval-2014 and 89.2\% accuracy on the IMDB movie review, both of which significantly outperform the comparative baseline model \cite{0.0}. Similarly, FGCN reaches 79.15\%, 47.04\%, 75.28\%, and 79.24\% accuracy on SST2, SemEval, MR, and M2SA, which is significantly better than the baseline models of CNN, RNN, and GCN, and especially outperforms the baseline model in ambiguous sentiment phrase processing \cite{0.1}. In the WSD, the Siamese model obtains F1 scores of 0.889 and 0.792 on ULS-WSD-18 and UAW-WSD-18, respectively \cite{0.2}. However, the above research advances face the common challenge of fast modeling of word polysemy \cite{0.3,0.4}. This problem might be prone to cause misunderstanding when words dynamically display multiple meanings in different contexts. For example, the word “bank” in “the bank was flooded” could be misclassified as a financial institution rather than a riverbank, distorting the results of analysis and compromising downstream applications such as risk assessment \cite{0.5}. In information retrieval, “apple” may refer to a fruit or a company, and failure to recognize the specific meaning can lead to irrelevant search results \cite{0.6}. Fortunately, the GA offers a promising path to address these limitations.

In recent years, GA has begun to receive attention in the field of NLP as an unstructured search quantum algorithm with the advantage of quadratic acceleration \cite{0.7}. In 2021, it was first used in question and answer tasks and optimized the time complexity from $O(P)$ of the classical algorithm to $O(\sqrt{P/Q})$, where $P$ is the total number of candidate answers and $Q$ is the number of correct answers, which significantly improved the answer retrieval efficiency \cite{0.8}. In 2024, quantum preprocessing by GA realizes about 50\% reduction in data processing time \cite{0.9}. In the same year, GA was fused with GPT with nearly 100\% accuracy in 6-qubit tests, and generalized to a 20-qubit systems with only 3 to 6 qubits of training data, far exceeding GPT-4o (45\%) \cite{0.10}. In 2025, GroverGPT-2 significantly outperformed other baseline models with search accuracy and fidelity consistently close to 1.0 in an Oracle-only input task simulating GA, and maintained highly stable performance at high qubit counts \cite{0.11}. Although the above case demonstrates the power of GA in NLP applications, the probability amplitude of the target state of GA decreases sharply when the number of target terms is unknown, which may affect its adaptability in other tasks \cite{0.12}.

To address the two major challenges of fast modeling of word polysemy and GA adaptive number of target states, the specific contributions of this paper are as follows:
\begin{itemize}
    \item  The QEPFE based on the adaptive GA is proposed for quantum modeling of polysemous words and fast search of word meanings.
    \item The QTP-Net is constructed based on the QEPFE and ERNIE \cite{3.4.0} for NLP tasks.
    \item The performance of the QTP-Net with frontier models is comprehensively compared and evaluated in SC and WSD tasks to verify its efficiency.
\end{itemize}

This paper is organized as follows: Section \ref{sec2} provides an overview of quantum computing foundations, GA and ERNIE as the theoretical basis of QEPFE. Section \ref{sec3} presents QEPFE in mathematical form and constructs the corresponding QTP-Net. Section \ref{sec4} reports the comparison results of QTP-Net in SC and WSD and its main findings. Finally, conclusions are drawn based on these findings.
\section{Preliminaries}\label{sec2}
This section provides a brief introduction to quantum computing foundations, the GA and the ERNIE.

\subsection{Quantum Computing Foundation}\label{Quantum gates}

Qubits and quantum gates are the figurative embodiment of quantum theory. In this context, a qubit 
\begin{equation}\label{qubit}
	\setlength{\abovedisplayskip}{3pt}\setlength{\belowdisplayskip}{3pt}
	|\psi \rangle =\alpha |0\rangle +\beta |1\rangle
\end{equation}
 is the smallest unit that carries information, where $|0\rangle ={{[1,0]}^{\text{T}}}$ and $|1\rangle ={{[0,1]}^{\text{T}}}$ denote the ground state and the excited state. $\alpha$ and $\beta$ are amplitudes satisfying $|\alpha|^2+|\beta|^2=1$ \cite{1.0}. In stark contrast to classical computation, \eqref{qubit} is in a superposition of $|0\rangle$ and $|1\rangle$, thereby conferring upon it the remarkable ability for exponential data representation. Moreover, the linear evolution of qubits is contingent upon quantum gates whose mathematical essence is the unitary matrices. The quantum gates used in this paper are shown in Tab. \ref{Notations}, including Pauli X gate, Pauli rotating X gate, Hadamard gate, controlled Y gate, multi-controlled Z gate.
 \begin{table*}[h]
	%
	\def\tablename{Tab.}
	\centering
	\caption{Quantum Gates}
	\label{Notations}
	\small
	\begin{tabular}{@{}lccc@{}}
		\toprule
		Name of Quantum Gate & Mathematical Notation & Matrix Representation & Symbol of Quantum Gate \\ \midrule
		
			Pauli X gate& $X$& $\left[ \begin{matrix}
			0   &1 \\ 
			1   &0  \\
		\end{matrix} \right]$
		
		&         
		\includegraphics[align=c,scale=0.126]{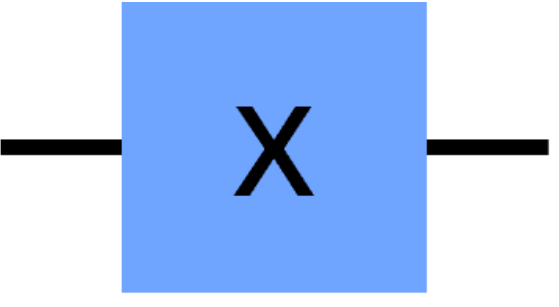} \\ \rule{-2pt}{22pt}
	
		Hadamard gate &    $H$ &$\frac{1}{\sqrt{2}}\left[ \begin{matrix}
		1 & 1  \\
		1 & -1  \\
	\end{matrix} \right]$
	&         \includegraphics[align=c,scale=0.1]{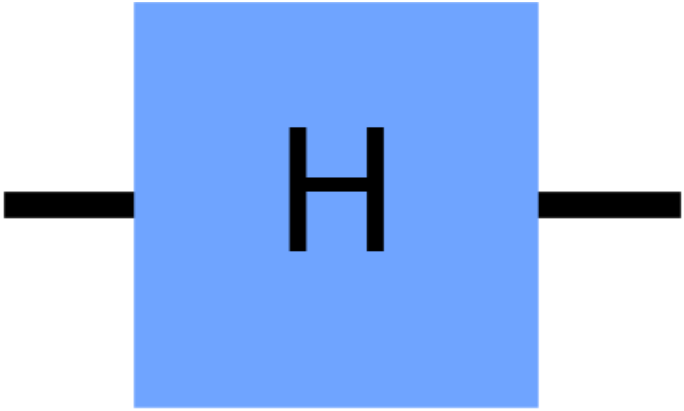}                                             	 \\  \rule{-2pt}{36pt}
	
	Multi-controlled X gate&    $MCX$ &$|0\rangle \langle 0|\otimes I+|1\rangle \langle 1|\otimes X$
	&         \includegraphics[align=c,scale=0.13]{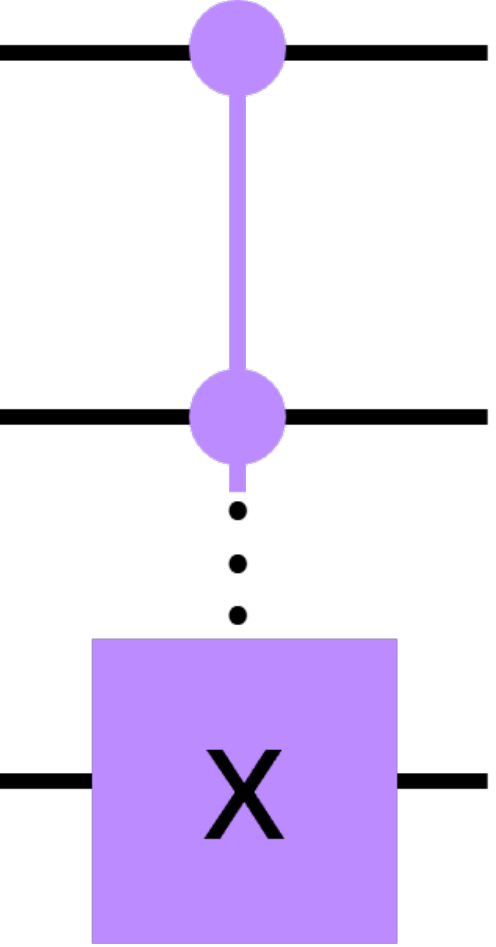}                                             	 \\ \rule{-2pt}{42pt}
		
		Multi-controlled Z gate &    $MCZ$ &$|0\rangle \langle 0|\otimes I+|1\rangle \langle 1|\otimes Z$
		&         \includegraphics[align=c,scale=0.13]{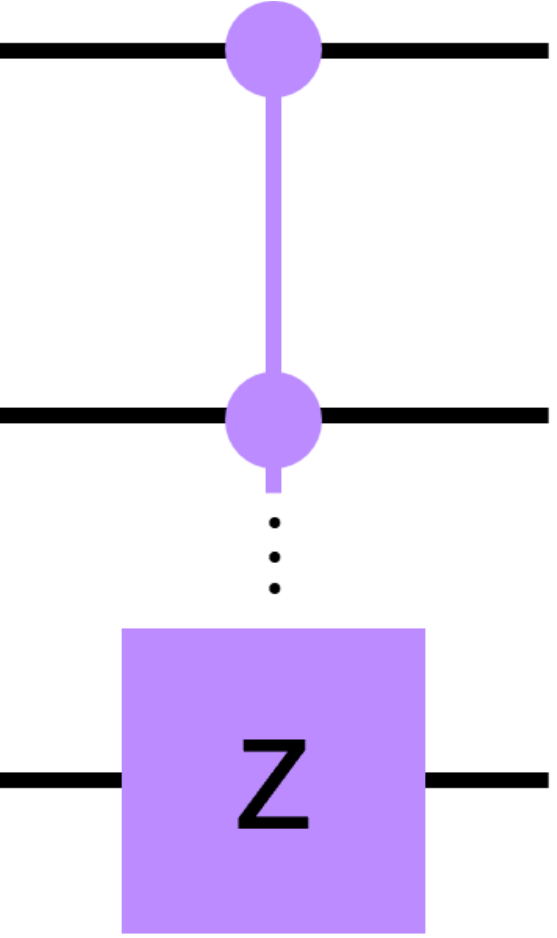}                                             	 \\  
		\bottomrule
	\end{tabular}%
\end{table*}

\subsection{Grover's Algorithm}
The GA \cite{0.7} iteratively locates a unique target state \( |\omega\rangle \) in an unsorted database of size \(N\) by amplifying its probability amplitude. It begins with the initial state 
\begin{equation}\label{init}
\setlength{\abovedisplayskip}{3pt}\setlength{\belowdisplayskip}{3pt}
|\mathbf{In}\rangle=\frac{1}{\sqrt{N}}\sum_{x=0}^{N-1}|x\rangle,
\end{equation}
where $|x\rangle$ is the computational basis state. An oracle \(Or\) then inverts the phase of the target state:
\begin{equation}\label{select}
\setlength{\abovedisplayskip}{3pt}\setlength{\belowdisplayskip}{3pt}
Or|x\rangle ={{(-1)}^{f(x)}}|x\rangle =\left\{ \begin{array}{*{35}{l}}
   -|x\rangle  & x=\omega   \\
   |x\rangle  & x\ne \omega   \\
\end{array} \right.,
\end{equation}
where \(f(x)=1\) when \(x=\omega\) and \(f(x)=0\) otherwise. Next, the diffusion operator
\begin{equation}\label{diff}
\setlength{\abovedisplayskip}{3pt}\setlength{\belowdisplayskip}{3pt}
D=2|\mathbf{In}\rangle\langle \mathbf{In}|-I,
\end{equation}
with \(I\) as the identity operator, reflects the state about the mean amplitude. Applying the Grover operator 
\begin{equation}\label{G1}
\setlength{\abovedisplayskip}{3pt}\setlength{\belowdisplayskip}{3pt}
    G=D\cdot Or
\end{equation}
approximately $\left\lfloor \tfrac{\pi }{4}\sqrt{M/N} \right\rfloor $ times maximizes the amplitude of the marked state \( |\omega\rangle \) when $M$ is known. Conversely, if $M$ is uncertain, the success probability may not be optimal.

\subsection{Enhanced Representation through kNowledge IntEgration 2.0}
The ERNIE 2.0 framework \cite{3.4.0} processes an input sequence $X = (x_1, x_2, \dots, x_m)$ by first computing its embedding $E = E_{\text{token}}(X) + E_{\text{sentence}}(X) + E_{\text{position}}(X) + E_{\text{task}}(X)$, where $E_{\text{token}}$, $E_{\text{sentence}}$, $E_{\text{position}}$, and $E_{\text{task}}$ represent the token, sentence, position, and task embedding functions, respectively, and then passes $E$ through a multi-layer Transformer encoder with $L$ layers, each applying self-attention and feed-forward transformations such that $H^0 = E$ and $H^l = \text{TransformerLayer}(H^{l-1})$ for $l = 1, \dots, L$, producing contextual representations $H^L$. During continual pre-training, the model incrementally constructs a series of tasks $\mathcal{T} = \{T_1, T_2, \dots, T_n\}$ and trains on subsets $\mathcal{T}_k \subseteq \mathcal{T}$ over $K$ stages, updating parameters $\theta$ from $\theta_{k-1}$ to $\theta_k$ by applying $n_{i,k}$ iterations per task $T_i \in \mathcal{T}_k$ using loss functions $\mathcal{L}_{T_i}(H^L)$, thereby encoding lexical, syntactic, and semantic information. For fine-tuning, the pre-trained model, initialized with $\theta_K$, is optimized with a downstream task $T_{\text{down}}$'s loss $\mathcal{L}_{T_{\text{down}}}(\theta)$ to adapt $H^L$ to specific language understanding applications.

\section{Technical Core and Framework of the Quantum Text Pre-training Network}\label{sec3}
In this section, a QEPFE combined with adaptive GA is proposed to realize multiple meaning encoding of words and the fast meaning speed diagnosis. Based on QEPFE and ERNIE, the QTP-Net framework is designed.

\subsection{Quantum Enhanced Pre-training Feature Embedding}
\begin{figure}[h]
    \centering
    \includegraphics[width=0.9\linewidth]{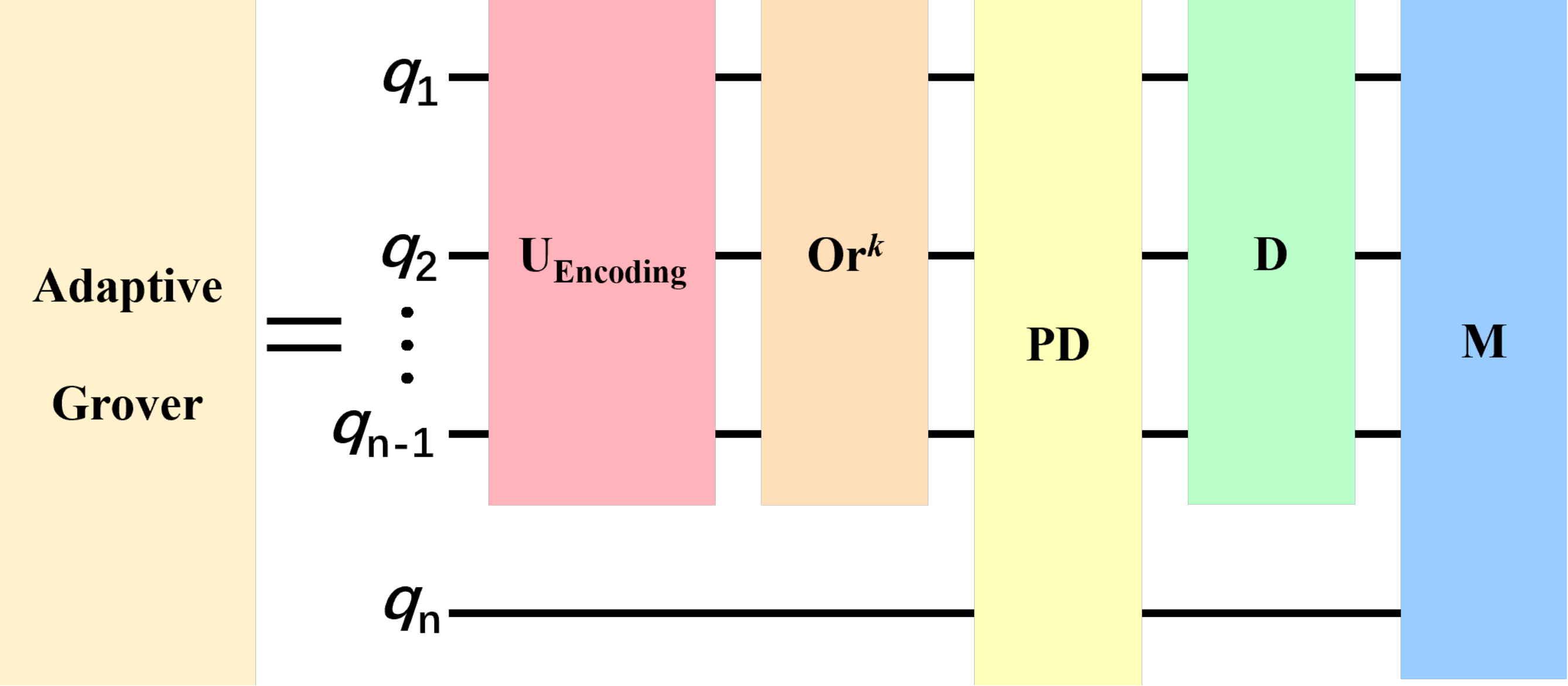}
    \caption{Adaptive GA framework}
    \label{GA}
\end{figure}

In NLP, polysemy, which includes semantic ambiguity and double meanings, is a core challenge in semantic modeling. At the same time, the number of word meanings may also be unknown in practical problems. To address this, an adaptive GA with $n$ qubits in Fig. \ref{GA} is defined as follows:
\begin{equation}\label{AGA}
\setlength{\abovedisplayskip}{3pt}\setlength{\belowdisplayskip}{3pt}
   \begin{aligned}[b]
   AGA&=AGAO\cdot {{U}_{\text{Encoding}}}|\mathbf{0}\rangle , \\ 
 & =D\cdot PD\cdot O{{r}^{k}}\cdot {{U}_{\text{Encoding}}}|\mathbf{0}\rangle .
\end{aligned}
\end{equation}
Obviously, 
\begin{equation}\label{G2}
\setlength{\abovedisplayskip}{3pt}\setlength{\belowdisplayskip}{3pt}
    AGAO:=D\cdot PD\cdot O{{r}^{k}},
\end{equation}
where $D$, $PD$, $Or^k$, $U_{\text{Encoding}}$ and $|\textbf{0}\rangle=|0^{\otimes n}\rangle$ are the adaptive diffusion operator, phase detection operator, oracle with unknown structure and encoding operator, and initial quantum state, respectively, where $\otimes$ is the symbol of the tensor. Notably, two key technologies have been developed to elegantly address the above problems. The first is the $Or^k$ adaptive algorithm, which is used to dynamically determine the number of semantics. The second is the $PD$, which automatically terminates when the amplitude of the target state reaches its maximum. The specific structures of all the above quantum operators and oracle diagnostic algorithms for unknown structures are presented next, respectively.
\subsubsection{Quantum Circuit Design of the Adaptive Grover's Algorithm}
\begin{figure}[h]
    \centering
    \includegraphics[width=1\linewidth]{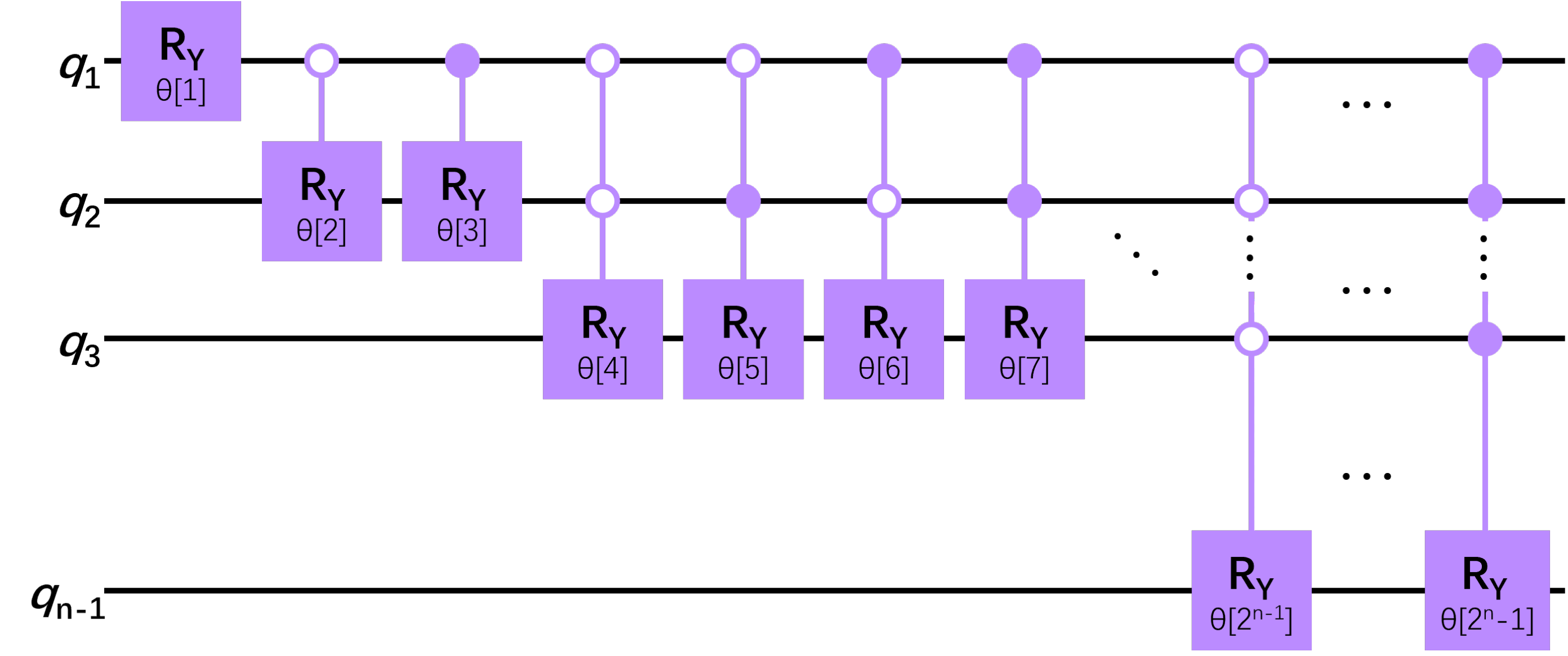}
    \caption{Amplitude Encoding Structure \cite{2.0}}
    \label{Ampli}
\end{figure}

\begin{figure}[h]
    \centering
    \includegraphics[width=0.6\linewidth]{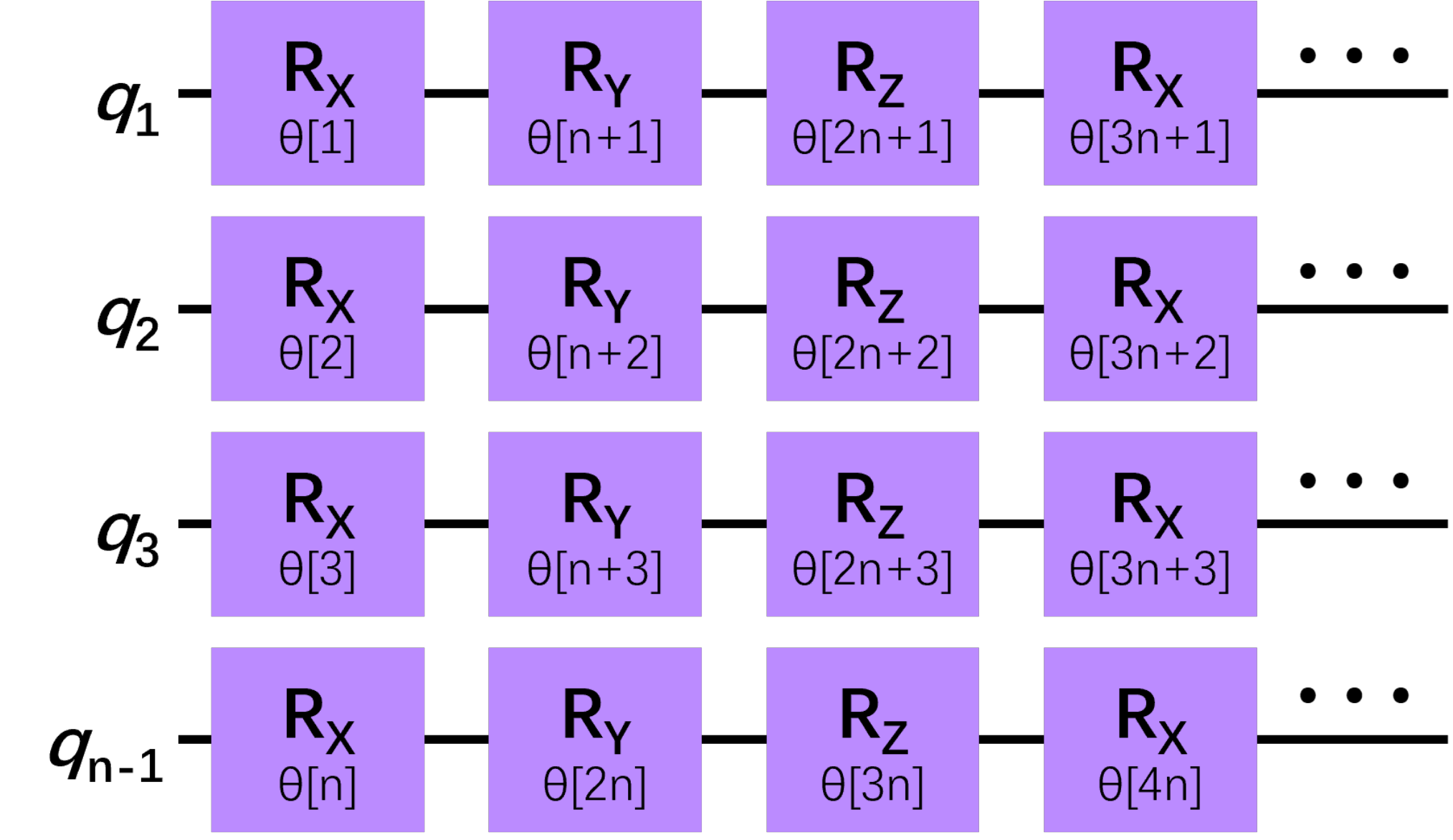}
    \caption{Angle Encoding Structure \cite{2.1}}
    \label{angle}
\end{figure}

\begin{figure}[h]
    \centering
    \includegraphics[width=0.5\linewidth]{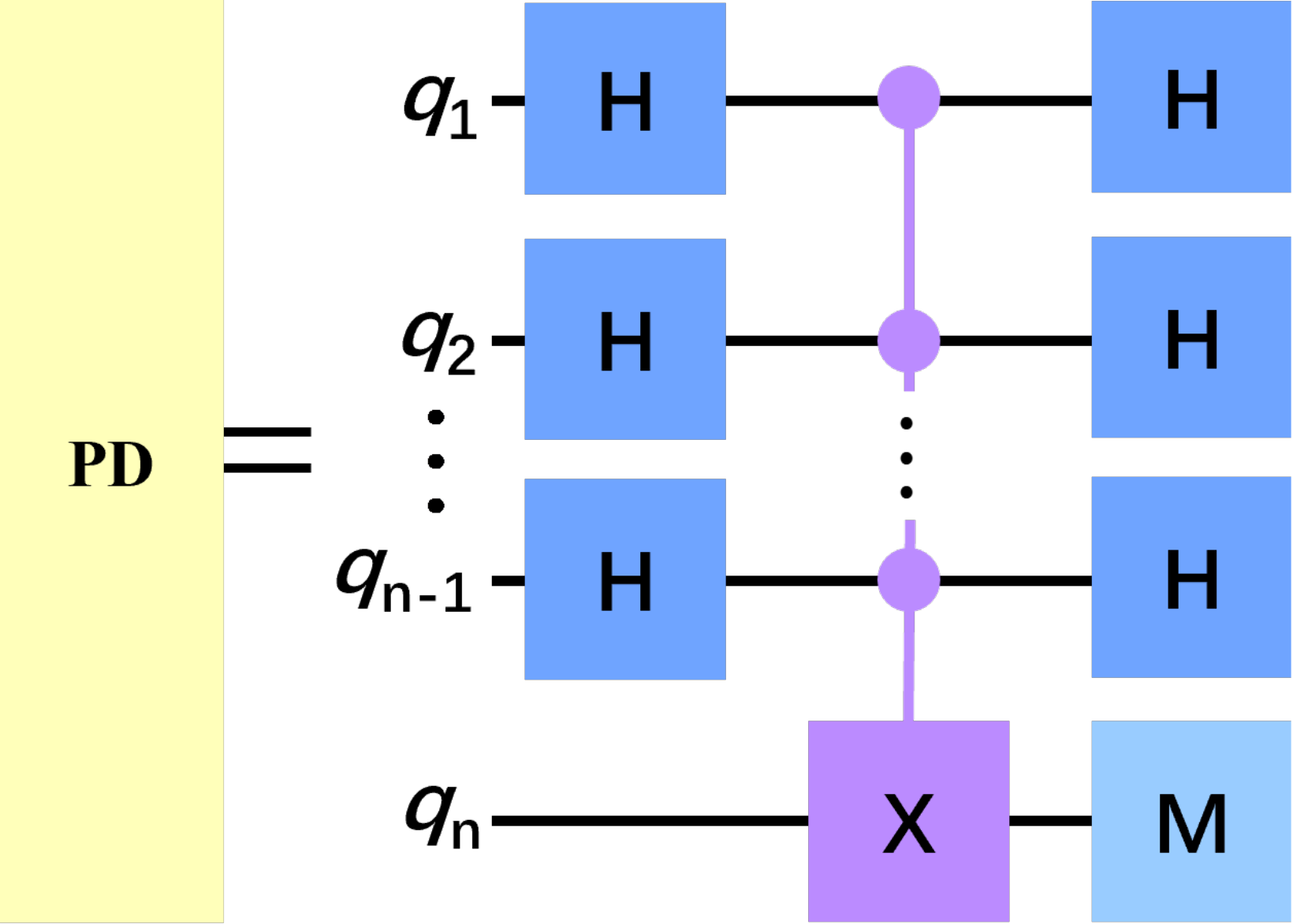}
    \caption{$PD$ Structure}
    \label{PD}
\end{figure}

\begin{figure}[h]
    \centering
    \includegraphics[width=0.8\linewidth]{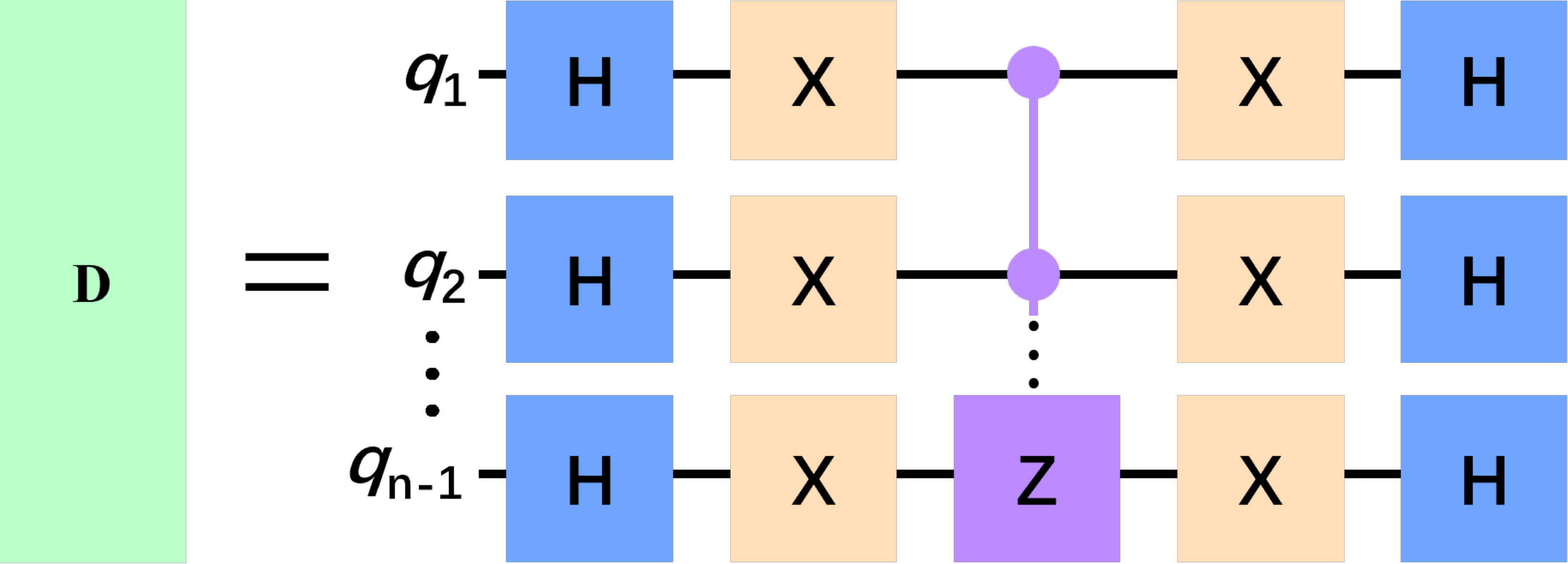}
    \caption{$D$ Structure}
    \label{D}
\end{figure}

First, $U_{\text{Encoding}}$ whose specific structure can be referred to the amplitude encoding \cite{2.0} in Fig. \ref{Ampli} or angle encoding \cite{2.1} in Fig. \ref{angle}, loads the word vector $W=[w_i]_{i=1}^m$ with ${{w}_{i}}\in \mathbb{R}$ into the Hilbert space $\mathcal{H}$:
\begin{equation}\label{encoding}
\setlength{\abovedisplayskip}{3pt}\setlength{\belowdisplayskip}{3pt}
W\xrightarrow{{{U}_{\text{Encoding}}}}|\mathcal{W}\rangle \in \mathcal{H}.
\end{equation}
In contrast to \eqref{select}, $Or^k$ needs to dynamically adjust its structure to discriminate the number of semantics depending on $|\mathcal{W}\rangle$. The structure of
\begin{equation}\label{PD_F}
\setlength{\abovedisplayskip}{3pt}\setlength{\belowdisplayskip}{3pt}
PD=(I-|s\rangle \langle s|)\otimes I+|s\rangle \langle s|\otimes X
\end{equation}
is as shown in Fig. \ref{PD}, where 
\begin{equation}\label{s}
\setlength{\abovedisplayskip}{3pt}\setlength{\belowdisplayskip}{3pt}
|s\rangle =|+\rangle^{\otimes (n-1)}={{\left( \frac{|0\rangle +|1\rangle }{\sqrt{2}} \right)}^{\otimes (n-1)}}.
\end{equation}
The principle of \eqref{PD_F} relies on the synergistic action of $H$ and $MCX$. First, \( H^{\otimes n} \) is applied to \( |\psi\rangle \) to convert it from the Z basis to the X basis, at which point the amplitude of \( |0\rangle^{\otimes n} \) is proportional to \(\phi_{++\cdots+} \). Next, $MCX$ uses the \( n \) qubits of \( |\psi\rangle \) as control bits and the last one as the target bit. When \(\phi_{++\cdots+}\leq 0 \), it flips the target bit. Finally, \( H^{\otimes n} \) is applied again to restore \( |\psi\rangle \) to the original basis, leaving it unchanged. Through this process, \eqref{PD_F} detects the phase sign and indicates the termination of iteration through the state change of the target bit. $D$, illustrated in Fig. \ref{D}, is consistent with the principle of \eqref{diff}. At this point, all structures are presented except for the structure of $Or^k$, which needs to be determined algorithmically.

\subsubsection{Structure-adaptive Algorithm for $Or^k$ }
When the number of marked items \(a\) is unknown, define
\begin{equation}
\setlength{\abovedisplayskip}{3pt}\setlength{\belowdisplayskip}{3pt}
\theta_a \;=\;\arcsin\sqrt{\frac{a}{N}},
\end{equation}
and let the number of Grover iterations \(k\) be drawn uniformly at random from \(\{1,\dots,m\}\).  After performing \(m\) randomized Grover iterations, the probability of measuring a marked element is
\begin{equation}
\setlength{\abovedisplayskip}{3pt}\setlength{\belowdisplayskip}{3pt}
P_m \;=\;\frac{1}{2}\;-\;\frac{\sin(4m\theta_a)}{4m\sin(2\theta_a)}.
\end{equation}
When
\begin{equation}
\setlength{\abovedisplayskip}{3pt}\setlength{\belowdisplayskip}{3pt}
m\;\ge\;\frac{1}{\sin(2\theta_a)},
\end{equation}
i.e.\ \(m\gtrsim\sqrt{N/a}\), one obtains \cite{2.2}
\begin{equation}
\setlength{\abovedisplayskip}{3pt}\setlength{\belowdisplayskip}{3pt}
P_m\;\ge\;\frac{1}{4}.
\end{equation}
Therefore, by starting from \(m=1\) and increasing \(m\) geometrically by a constant factor \(\lambda>1\) (commonly \(\lambda=6/5\)), one finds a marked element with constant success probability in \(O(\sqrt{N/a})\) total iterations. The pseudo code is as illustrated in Alg. \ref{alg1}:
\begin{algorithm}[ht]
\caption{Quantum Search with Unknown Number of Solutions \cite{2.2}}
\label{alg1}
\begin{algorithmic}[1]
\Require  \(Or^k\) such that \(f(x)=1\iff x\in A\); growth factor \(\lambda=6/5\).
\Ensure An element \(x\in A\).
\State \(m \gets 1\).
\While{\(m \le \sqrt{N}\)}
  \State Sample \(k\) uniformly from \(\{1,\dots,m\}\).
  \State Prepare \eqref{encoding}.
  \For{\(i=1\) to \(k\)}
    \State Apply one Grover iteration \eqref{G2}.
  \EndFor
  \State Measure to obtain \(x\).
  \If{\(x\in A\)}
    \State \Return \(x\).
  \Else
    \State \(m \gets \lambda\,m\).
  \EndIf
\EndWhile
\end{algorithmic}
\end{algorithm}

\subsection{Quantum Text Pre-training Network}
The QTP-Net framework, as shown in Fig. \ref{QTPNET}, integrates QEPFE and ERNIE for efficient semantic identification and feature extraction. The input token sequence \(X\) is processed through two parallel branches:
\begin{figure}[h]
    \centering
    \includegraphics[width=0.5\linewidth]{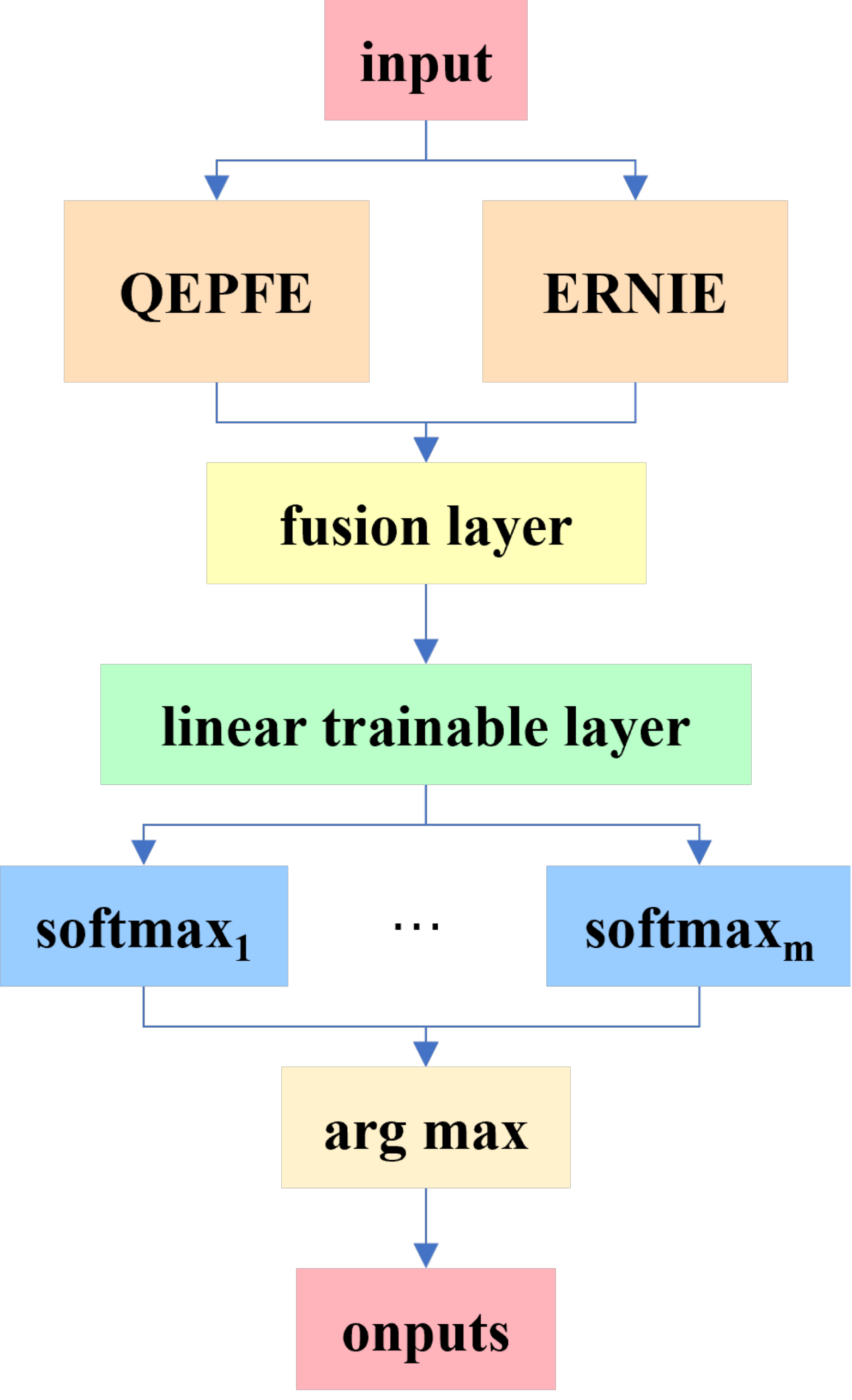}
    \caption{Framework of QTP-Net}
    \label{QTPNET}
\end{figure}
(1) \textbf{QEPFE Branch}: \(X\) is embedded into QEPFE, undergoes \(k\) iterations of \eqref{AGA}, and the probability distribution
\begin{equation}
\setlength{\abovedisplayskip}{3pt}\setlength{\belowdisplayskip}{3pt}
p(x) = ||\langle \Psi_0 | AGA \rangle||^2
\end{equation}
is collected as the feature vector \(\mathbf{p} = [p(x)]_{x=0}^{N-1}\).
(2) \textbf{ERNIE Branch}: The high-level semantic embedding
\begin{equation}
\setlength{\abovedisplayskip}{3pt}\setlength{\belowdisplayskip}{3pt}
\mathbf{h} = \mathrm{ERNIE}(X)
\end{equation}
is computed.
The outputs are fused by concatenation:
\begin{equation}
\setlength{\abovedisplayskip}{3pt}\setlength{\belowdisplayskip}{3pt}
\mathbf{z} = \mathbf{p} \oplus \mathbf{h},
\end{equation}
and passed through a linear trainable layer followed by softmax to produce the prediction
\begin{equation}
\setlength{\abovedisplayskip}{3pt}\setlength{\belowdisplayskip}{3pt}
\hat{y} = \arg\max \text{softmax}(W\mathbf{z} + b).
\end{equation}
During training, ERNIE parameters are frozen, and only the weights of the linear trainable layer are optimized using the cross-entropy loss \(\mathcal{L} = -\sum_i y_i \log \hat{y}_i\), with a classical optimizer (e.g., RAdam) for backpropagation. 
\begin{algorithm}[ht]
\caption{Training Algorithm for QTP-Net}
\label{alg:QTPNet}
\begin{algorithmic}[1]
\Require Token sequence $X$, label $y$, max epochs $E$
\Ensure Predicted label $\hat{y}$
\State Freeze ERNIE parameters; initialize QEPFE and fusion layer weights
\For{epoch = 1 to $E$}
  \State \textbf{Quantum Branch:}
  \State \quad Prepare initial state $|\Psi_0\rangle$
  \State \quad Repeat QEPFE approximately $O(\sqrt{N})$ times to obtain $AGA$
  \State \quad Measure to get probability vector $\mathbf{p}$
  \State \textbf{ERNIE Branch:}
  \State \quad Compute semantic embedding $\mathbf{h} \gets \mathrm{ERNIE}(X)$
  \State \textbf{Fusion \& Prediction:}
  \State \quad Concatenate $\mathbf{z} \gets [\mathbf{p} \,\|\, \mathbf{h}]$
  \State \quad Compute logits $\ell \gets W\mathbf{z} + b$
  \State \quad Predict $\hat{y} \gets \arg\max \text{softmax}(\ell)$
  \State \textbf{Loss \& Update:}
  \State \quad Compute $\mathcal{L} \gets -\sum_{i} y_i \log \hat{y}_i$
  \State \quad Backpropagate and update fusion weights via RAdam
  \If{converged}
    \State \textbf{break}
  \EndIf
\EndFor
\State \Return $\hat{y}$
\end{algorithmic}
\end{algorithm}

\section{Experiments and Data Analysis}\label{sec4}
In this section, the performance of QTP-Net is comprehensively evaluated on PennyLane \cite{3.0} and PyTorch \cite{3.1} platforms. Specifically, two categories of experiments are conducted:

(1) SC: QTP-Net, four classical models (i.e. VanillaGRU \cite{3.2}, BiLSTM \cite{3.3}, TextCNN \cite{3.4} and ERNIE \cite{3.4.0}) and three quantum-inspired models (i.e. CE-Mix \cite{3.4.1}, CNN-Complex-order \cite{3.4.2} and TextTN \cite{3.4.3}) are tested and compared on the Customer Review (CP) \cite{3.5}, Multi-Perspective Question Answering (MPQA) \cite{3.6}, Movie Review (MR) \cite{3.7}, Stanford Sentiment Treebank (SST) \cite{3.8}, Subjectivity (SUBJ) \cite{3.9}, and Chinese Sentiment Corpus (ChnSentiCorp) \cite{3.10} datasets, respectively, highlighting the sentiment classification advantage of QTP-Net.

(2) WSD: QTP-Net, BERT \cite{3.11}, ERNIE, EWISE \cite{3.12}, GLU \cite{3.13}, SyntagRank \cite{3.14}, SREF \cite{3.15}, Generationary \cite{3.16}, and GlossBERT \cite{3.17} are tested and compared on the SemCor \cite{3.17.1}, Senseval-2 (SE2) \cite{3.18}, Senseval-3 (SE3) \cite{3.19}, SemEval-2007 (SE07) \cite{3.20}, SemEval-2013 (SE13) \cite{3.21} and SemEval-2015 (SE15) \cite{3.22} datasets, respectively, emphasizing the word sense disambiguation advantage of QTP-Net. 

\subsection{Dataset and Configurations}
\subsubsection{Datasets}
Six datasets, including CR, MPQA, MR, SST, SUBJ and ChnSentiCorp, are used in the SC task. They contain customer product reviews, movie review sentences, opinion spans extracted from news articles, provide phrase and sentence level annotations in syntactic parsing structures, subjective expressions and objective statements and collect Chinese hotel reviews respectively, all suitable for sentiment binary categorization, and each benchmark contains text snippets labeled according to positive or negative polarity. Their data sizes are approximately 4k, 11k, 11k, 70k, 10k, and 7k respectively. The WSD baseline exploits a richly annotated SemCor training set alongside five standardized evaluation benchmarks. SemCor provides 87002 noun, 88334 verb, 31753 adjective and 18947 adverb instances (226036 tokens in total) drawn from balanced news and narrative text, offering comprehensive coverage across parts of speech for supervised sense learning. Its test suites include Senseval‐2 (SE2), comprising 1066 nouns, 517 verbs, 445 adjectives and 254 adverbs (2282 total); Senseval‐3 (SE3), with 900 nouns, 588 verbs, 350 adjectives and 12 adverbs (1,850 total); SemEval‐2007 (SE07), focusing on 159 nouns and 296 verbs (455 total); SemEval‐2013 (SE13), targeting 1644 noun instances; and SemEval‐2015 (SE15), containing 531 nouns, 251 verbs, 160 adjectives and 80 adverbs (1022 total). Together, these datasets span multiple lexical categories and domain conditions, forming a rigorous foundation for evaluating both coarse‐ and fine‐grained WSD systems.

\subsubsection{Configurations}
For SC, QTP-Net has a learning rate of 0.00001, epochs of 5, and a batch size of 32. For WSD, QTP-Net has a learning rate of 0.0003, epochs of 30, and a batch size of 50, and is trained on the SemCor corpus and then evaluated on five additional datasets. For this experiment, the following four main indicators are considered:
\begin{equation}\label{acc}
    \setlength{\abovedisplayskip}{3pt}\setlength{\belowdisplayskip}{3pt}\text{accuracy} = \frac{\text{TP + TN}}{\text{TP + TN + FP + FN}}.
\end{equation}
\begin{equation}\label{pre}
    \setlength{\abovedisplayskip}{3pt}\setlength{\belowdisplayskip}{3pt}\text{precision} = \frac{\text{TP}}{\text{TP + FP}}.
\end{equation}
\begin{equation}\label{rec}
    \setlength{\abovedisplayskip}{3pt}\setlength{\belowdisplayskip}{3pt}\text{recall} = \frac{\text{TP}}{\text{TP + FN}}. 
\end{equation}
\begin{equation}\label{f1}
    \setlength{\abovedisplayskip}{3pt}\setlength{\belowdisplayskip}{3pt}\text{F1 score} = 2 \times \frac{\text{precision} \times \text{recall}}{\text{precision} + \text{recall}}.
\end{equation}
Here, TP, TN, FP and FN represent true positives,  true negatives, false positives, false negatives. 
\begin{figure*}[h!]
    \centering
    \includegraphics[width=1\textwidth]{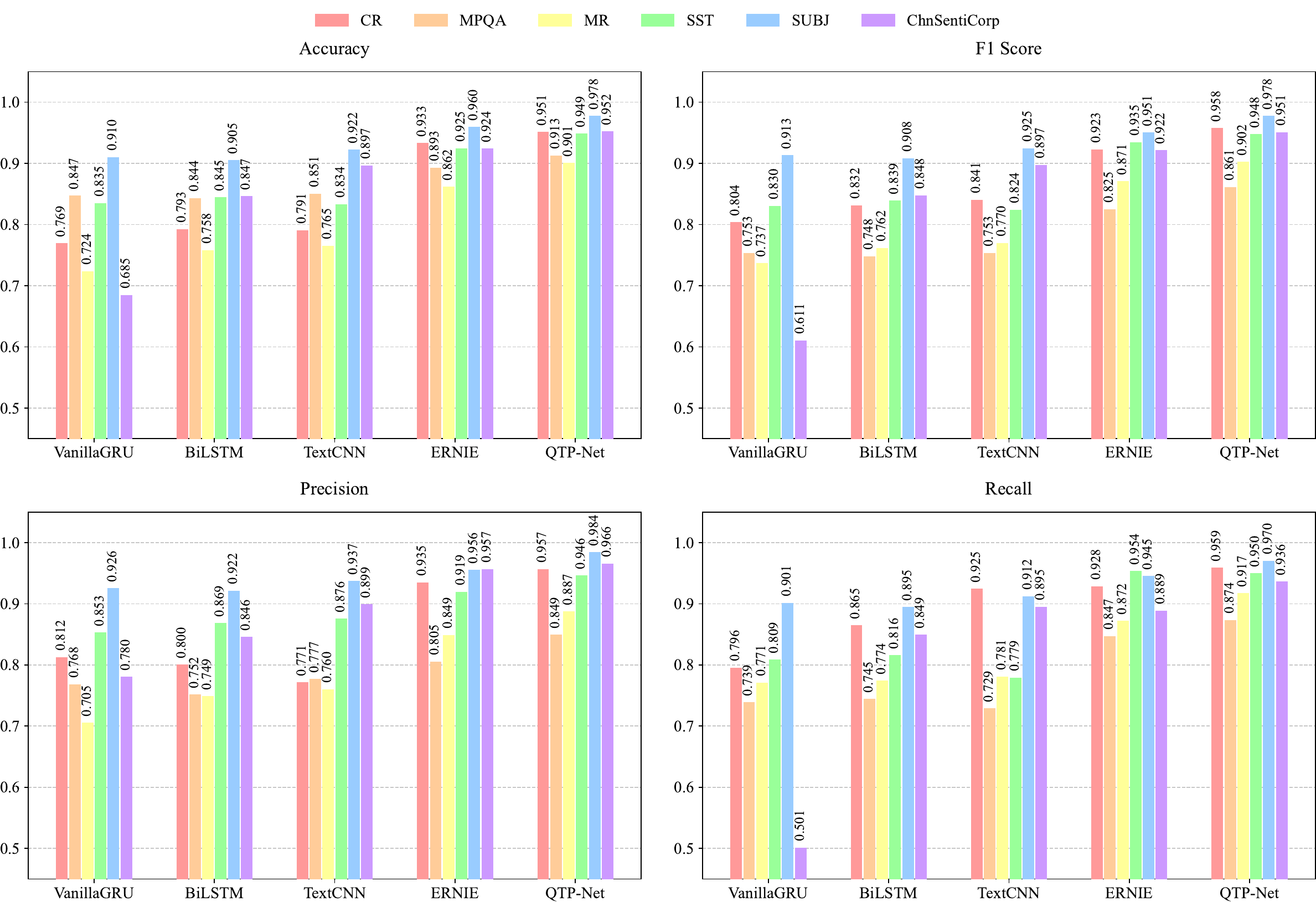}
    \caption{Comparative results of sentiment classification datasets}
    \label{res1}
    \centering
    \includegraphics[width=1\linewidth]{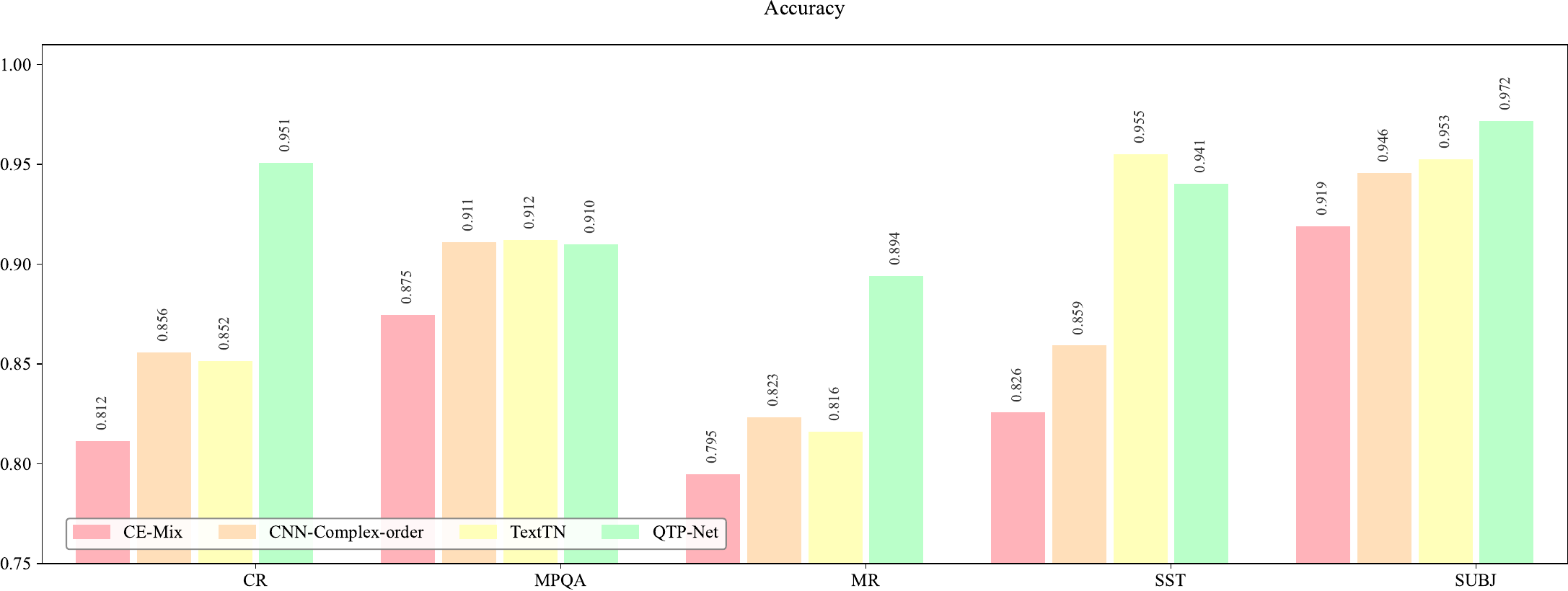}
    \caption{Sentiment classification accuracy compared to quantum-inspired models}
    \label{res2}
\end{figure*}
\begin{figure*}[h]
    \centering
    \includegraphics[width=1\linewidth]{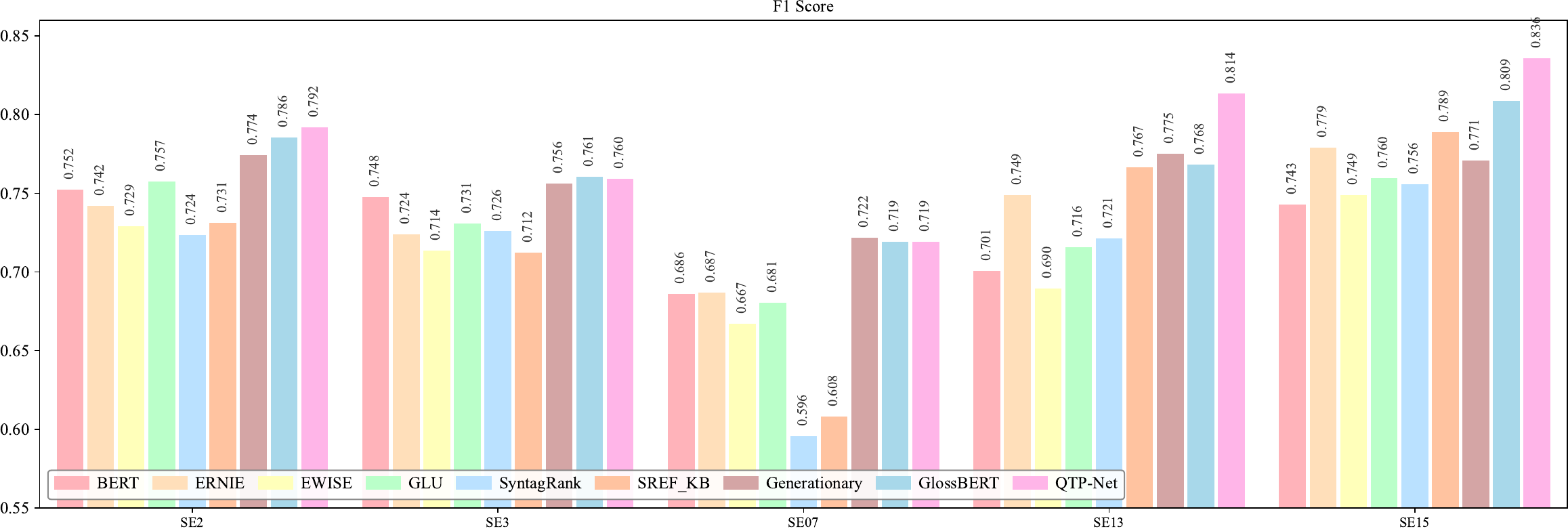}
    \caption{Experimental results of F1 score trained on SemCor and evaluated on the benchmark WSD dataset}
    \label{res3}
\end{figure*}

\subsection{Sentiment Classification}

The results of comparing QTP-Net with classical models \cite{3.2,3.3,3.4,3.4.0} are shown in Fig. \ref{res1}. The four subfigures portray indicators \eqref{acc} to \eqref{f1}. Their horizontal coordinates refer to the model, and the six colored bars indicate datasets \cite{3.5,3.6,3.7,3.8,3.9,3.10}. Each colored bar is labeled with the corresponding value at the top. According to Fig. \ref{res1}, the following conclusions can be drawn.
(1) From the accuracy perspective QTP-Net outperforms every baseline on all six datasets with an average improvement of 0.024 ($\pm$0.007) and gains ranging from 0.018 on MR to 0.039 on CR.
(2) In terms of F1 score QTP-Net yields an average lift of 0.029 over the second-best model, with the smallest margin of 0.013 on MPQA and the largest of 0.036 on CR.
(3) Looking at precision QTP-Net improves by an average of 0.028 compared to competitors, with gains between 0.009 on SUBJ and 0.044 on CR.
(4) For recall QTP-Net shows an average increase of 0.028 but experiences a slight 0.004 drop on SST relative to ERNIE, while all other datasets exhibit positive recall gains up to 0.045 on MPQA.
(5) QTP-Net maintains highly consistent performance across its four metrics on each dataset, with internal ranges of at most 0.008 on CR and 0.004 on SST, reflecting balanced improvements rather than trade-offs.
(6) Across datasets QTP-Net achieves its highest accuracy and F1 on SUBJ (0.978 and 0.978 respectively) and its lowest on ChnSentiCorp (0.952 and 0.951), indicating robust generalization even on Chinese sentiment data.

The results of the comparison between QTP-Net and three quantum-inspired models \cite{3.4.1,3.4.2,3.4.3} are shown in Fig. \ref{res2}. In Fig. \ref{res2}, the horizontal coordinates indicate the dataset, and the four colored bars denote the different models. This time, only accuracy is analyzed. According to Fig. \ref{res2}, the following conclusions can be obtained: (1) QTP-Net achieves an average accuracy of 0.934 across five datasets, exceeding TextTN by 0.036 and thereby suggesting a comparatively stronger ability to generalize.
(2) On CR, MR and SUBJ, QTP-Net attains accuracies of 0.951, 0.894 and 0.972 with margins of 0.095, 0.071 and 0.026 over the next best model, a difference that appears most pronounced in these scenarios.
(3) Accuracy scores of 0.910 on MPQA and 0.941 on SST fall short of TextTN by 0.002 and 0.015, indicating that classification of brief emotional passages and longer sentiment texts may warrant further attention.
(4) The accuracy spread for QTP-Net across five tasks is only 0.077, considerably lower than the spreads observed for CE-Mix at 0.125, CNN-Complex-order at 0.122 and TextTN at 0.139, reflecting a high degree of performance consistency.
(5) With an accuracy of 0.972 on SUBJ, QTP-Net displays a robust capacity for identifying nuanced sentiment features.
(6) An accuracy gain of 0.095 on CR highlights QTP-Net ability to model domain vocabulary and contextual dependencies with greater effectiveness.

\subsection{Word Sense Disambiguation}
The F1 scores of QTP-Net and the classical modelers \cite{3.4.0,3.11,3.12,3.13,3.14,3.15,3.16,3.17} on the WSD task are shown in Fig. \ref{res3}. In Fig. \ref{res3}, the horizontal axis indicates different test datasets \cite{3.18,3.19,3.20,3.21,3.22} and different colored bars to refer to the models. Based on Fig. \ref{res3}, the following conclusions can be found. (1) QTP-Net attains the highest mean F1 score of 0.784 across five evaluation sets, exceeding the next leading approach GlossBERT by 0.016, which suggests a broadly superior average effectiveness.
(2) On SE2, SE13 and SE15, QTP-Net records F1 values of 0.792, 0.814 and 0.836, respectively, with margins of 0.006, 0.038 and 0.027 over the second‐ranked method, indicating a more pronounced advantage in these contexts.
(3) Performance on SE3 and SE07 yields F1 scores of 0.760 and 0.719 for QTP-Net, falling short by approximately 0.001 and 0.003 relative to GlossBERT and Generationary, which points to modest opportunities for refinement on these subsets.
(4) The span between the highest and lowest F1 scores for QTP-Net amounts to 0.117, situating its task-to-task variation below that of SyntagRank and SREF yet above that of Generationary and BERT, reflecting moderate stability in cross‐domain performance.
(5) Comparison of average F1 by model places QTP-Net at the forefront, followed by GlossBERT at 0.769 and Generationary at 0.760, thereby indicating consistent advantages over established transformer and knowledge‐enhanced baselines.
(6) The largest relative improvement for QTP-Net appears on SE13 where the F1 gap to the nearest contender reaches 0.038, underscoring its capacity to capture complex relational patterns in this dataset.

\section{Conclusions}\label{sec5}
This work presents QTP-Net, an innovative framework that leverages quantum computing to address polysemy in NLP, a persistent challenge in semantic modeling. By introducing the QEPFE, we enable the representation of multiple word meanings in quantum superposition states, enhancing adaptability and semantic precision. Integrated with ERNIE, QTP-Net delivers significant improvements in SC and WSD tasks. Experimental results show QTP-Net surpasses classical and quantum-inspired models, achieving average gains of 2.4\% in accuracy and 2.9\% in F1 score across six SC datasets, and a leading F1 score of 0.784 in WSD, exceeding GlossBERT by 1.6\%. These outcomes validate the efficacy of quantum principles in NLP and establish QTP-Net as a robust solution for real-world language processing. Future research will focus on refining quantum circuit designs and extending QTP-Net to broader NLP applications.

\end{document}